\documentclass[comsoc, conference,10pt]{IEEEtran}
\usepackage{cite}
\usepackage{amsmath,amssymb,amsfonts}
\usepackage{multirow}
\usepackage{graphicx}
\usepackage{textcomp}
\usepackage{xcolor}
\usepackage{svg}
\usepackage{cleveref}
\usepackage{centernot}
\usepackage{booktabs}
\usepackage{url}

\newtheorem{example}{Example}

\usepackage[T1]{fontenc}

\usepackage{xcolor,colortbl}
\definecolor{celadon}{rgb}{0.67, 0.88, 0.69}
\definecolor{celestialblue}{rgb}{0.29, 0.59, 0.82}
\definecolor{cream}{rgb}{1.0, 0.99, 0.82}
\newcolumntype{n}{>{\columncolor{cream}}r}
\newcolumntype{f}{>{\columncolor{celadon}}r}
\newcolumntype{g}{>{\columncolor{celestialblue}}r}

\newcommand\opt[1]{\textcolor{blue}{\textbf{#1}}}

\usepackage{xcolor, soul}

\definecolor{bananamania}{rgb}{0.98, 0.91, 0.71}
\sethlcolor{bananamania}

\usepackage[titlenumbered,ruled,linesnumbered,noend]{algorithm2e}
\usepackage{subcaption}

\usepackage{MnSymbol,graphicx}
\newcommand\Bigcircle{\raisebox{-0.5mm}{\scalebox{2.7}{$\circ$}}}
\newcommand\Bigxor{\raisebox{-0.5mm}{\scalebox{1.7}{$\oplus$}}}

\usepackage[framemethod=tikz]{mdframed}
\usepackage{lipsum}

\title{\huge AnySyn: A Cost-Generic Logic Synthesis Framework\\ with Customizable Cost Functions}

\author{
    \IEEEauthorblockN{
        Hanyu Wang\\
        \textit{ETH, Zurich, Swizterland}
    }
\and
    \IEEEauthorblockN{
        Siang-Yun Lee\\
        \textit{EPFL, Lausanne, Swizterland}
    }
\and
    \IEEEauthorblockN{
        Giovanni De Micheli\\
        \textit{EPFL, Lausanne, Swizterland}
    }
}

\begin{document}

\makeatletter
\patchcmd{\@maketitle}
  {\addvspace{0.5\baselineskip}\egroup}
  {\addvspace{-0.8\baselineskip}\egroup}
  {}
  {}
\makeatother
\maketitle

\begin{abstract}
Modern technology-independent logic synthesis has been developed to optimize for the size and depth of AND-Inverter Graphs (AIGs) as a proxy of CMOS circuit area and delay. However, for non-CMOS-based emerging technologies, AIG size and depth may not be good cost estimations. Dedicated algorithms optimizing for more complex cost functions have been proven effective for their specific target applications yet require time and experts in both logic synthesis and the targeted technology to develop. In this work, we propose AnySyn, a cost-generic optimization framework for agile experimentation and prototyping of various customized cost functions before investing in developing specialized algorithms. Experimental results show that AnySyn outperforms non-specialized size and depth optimization algorithms by 14\% and 19\% on average and achieves comparable results to specialized algorithms within acceptable CPU time. 
\end{abstract}

\section{Introduction}
Logic synthesis has been developed around NAND-based CMOS technologies. Since the 2000s, \emph{AND-Inverter Graphs}~(AIGs) have been heavily used as the underlying multi-level logic network representation in scalable technology-independent logic synthesis~\cite{Mishchenko06}. Most modern logic optimization algorithms target minimizing AIG size (number of AND2 gates) or depth (length of critical path) because these cost functions are simple and often correlate with circuit area and delay, respectively. In the design of optimization algorithms, the targeted cost function plays an important role in guiding the direction of heuristic optimization and choosing among various optimization choices. 

However, for non-CMOS-based technologies and applications not based on NAND gates, optimizing for AIG size or depth does not always lead to the best Quality of Results (QoR). More complex cost functions are more often used recently to provide better QoR in terms of the actual cost of concern. 
For example, in cryptography and security applications, XOR gates are preferred over AND gates~\cite{TestaSAM19}, and in quantum circuits, XOR gates are much cheaper than AND gates~\cite{meuli2022xor}, thus \emph{Multiplicative Complexity}~(MC)~\cite{Schnorr88}, or the AND-count or AND-depth in an \emph{XOR-AND-Inverter Graph}~(XAG), is often used instead of AIG size or depth. As another example, the \emph{Factored Form Literal Cost}~(FFLC) has been shown to correlate better to technology-mapped results than AIG size~\cite{Calvino23}.
Specialized algorithms have been proposed, targeting these non-conventional cost functions, and have successfully achieved better QoR~\cite{TestaSAM19,Calvino23}. 

Experiments are needed to show the effectiveness of newly-proposed cost functions. Simple technology-independent representations such as AIGs and XAGs must be used to keep algorithms scalable. On the other hand, cost functions defined over such representations are always estimations of the final QoR metric. There could be multiple ways to define cost functions for the same target QoR metric. Without experimentation, it is unclear whether optimizing for a certain cost function indeed improves QoR. However, it may take weeks or months for an engineer to develop a specialized algorithm targeting a newly-defined cost function. Thus, a platform for quickly experimenting and prototyping different possible cost functions is in need.

In this paper, we propose AnySyn, a cost-generic logic synthesis framework that optimizes a technology-independent logic representation according to user-defined cost functions. The main challenge in supporting customizable cost functions is evaluating the change in cost that a local optimization choice would make in the global context. AnySyn performs such evaluation flexibly and efficiently by propagating global context and isolating local cost contributions. AnySyn optimizes in an algorithmic flow similar to Boolean resubstitution, but uses a cost-generic synthesis engine with crafted heuristics to enhance runtime efficiency. Supported by experimental results, we show that (1) AnySyn is compatible with a wide range of practical cost functions and outperforms non-specialized algorithms in nine example cost functions; (2) AnySyn, which can be specialized and tested in a day, achieves comparable QoR to a human-designed specialized algorithm, which may cost weeks to develop; (3) AnySyn maintains the same scalability as specialized algorithms with only a constant-ratio runtime overhead as a tradeoff of being generic.

\begin{figure*}[tb]
\captionsetup[subfigure]{justification=centering}
        \centering
    \begin{subfigure}[b]{.29\linewidth}
    \centering
        \includegraphics[width=.95\linewidth]{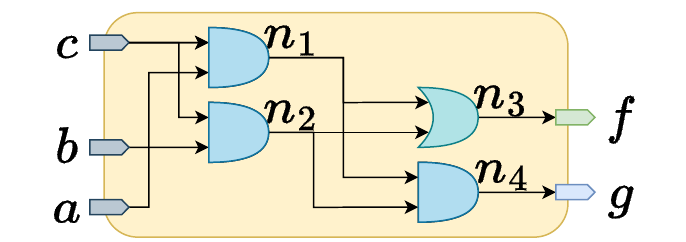}
        \caption{Logic network $N_A$}
        \label{fig:example-network-1}
    \end{subfigure}
    \begin{subfigure}[b]{.29\linewidth}
    \centering
        \includegraphics[width=.95\linewidth]{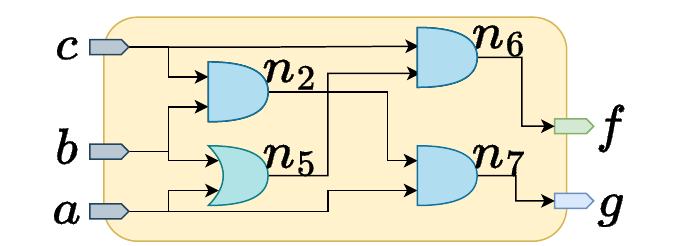}
        \caption{Logic network $N_B$}
        \label{fig:example-network-2}
    \end{subfigure}
    \begin{subfigure}[b]{.36\linewidth}
    \centering
        \includegraphics[width=.95\linewidth]{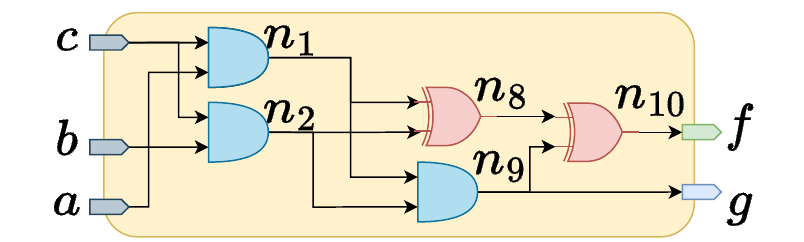}
        \caption{Logic network $N_C$}
        \label{fig:example-network-3}
    \end{subfigure}
    \caption{\small Example of three functionally-equivalent networks.}
    \label{fig:motivational-example}
\end{figure*}
\section{Background}\label{sec:background}

\subsection{Logic Networks}
\emph{Logic networks} are technology-independent representations of gate-level digital circuits. A logic network is a directed acyclic graph whose vertices, referred to as \emph{nodes}, represent logic gates or \emph{primary inputs}~(PIs), and edges represent wires. Incoming edges of a node $n$ are referred to as the \emph{fanins} of $n$, and the set of fanins is represented as $\delta^-(n)$. Similarly, the outgoing edges of a $n$ are called the \emph{ fanouts} of $n$. Examples of logic networks include AIGs~\cite{KuehlmannPKG02AIG} and \emph{XOR-AND-Inverter Graphs}~(XAGs)~\cite{HalecekFS17XAG}.

A \emph{cut} $C$ in a network is a tuple $(r, L)$ of a \emph{root} node $r$ and a set of \emph{leaf} nodes $L$, such that all paths from PIs to $r$ pass through a leaf in $L$. A \emph{reconvergence-driven cut} of node $r$ can be computed by heuristically picking one leaf $n$ to be replaced by its fanins (to \emph{expand} on $n$)~\cite{Mishchenko06}.

A \emph{cone} is the set of nodes on any path between a node $n$ and any leaf node in a cut rooted at $n$. The \emph{transitive-fanin cone}~(TFI) of a node $n$ is the cone between $n$ and the set of PIs. A \emph{fanout-free cone}~(FFC) of a node $n$ is a cone between $n$ and a cut $C$, where all paths from any leaf in $C$ to any PO pass through $n$. The \emph{maximum fanout-free cone}~(MFFC) of a node $n$ is the maximum-sized FFC of $n$. The MFFC of a node can be identified by recursively dereferencing and referencing the TFI of $n$~\cite{MishchenkoCB06}.

\subsection{Boolean Resubstitution}\label{subsec:resub}


\emph{Boolean Resubstitution}~\cite{Mishchenko06,RienerTASM18} is a scalable Boolean method in technology-independent logic optimization. Although there are different resynthesis techniques~\cite{LeeM23}, a resubstitution algorithm generally works by choosing a root node $n$, computing a reconvergence-driven cut $C$, and constructing a \emph{window} including the cone between $n$ and $C$ and nodes outside the cone but supported by $C$. The algorithm then resynthesizes $n$ using the local function of $n$ as the target and the local functions of the nodes in the window but not in the MFFC of $n$ as \emph{divisors}. Finally, the algorithm replaces $n$ with the resynthesized \emph{dependency circuit}.

\section{Motivating Example}\label{sec:motivation}
We use the three circuits in \Cref{fig:motivational-example} to show the limitations of existing logic synthesis algorithms and motivate our work. \Cref{fig:example-network-1}, \Cref{fig:example-network-2}, \Cref{fig:example-network-3} depict three logic networks where $a$, $b$, $c$ represent inputs, $f$ and $g$ represent outputs, and $n_1$ to $n_{10}$ are nodes that propagate logic functions. The truth table below express $n_1$ to $n_{10}$ as functions of $a$, $b$ and $c$.

\begin{table}[h]
\small
\tabcolsep = 5pt
    \centering
    \begin{tabular}{|rrr|rrfgrfgrgf|}
    \hline
    $a$ &$b$ &$c$ &$n_1$ &$n_2$ &$n_3$ &$n_4$ &$n_5$ & $n_6$ &$n_7$ &$n_8$ &$n_9$ &$n_{10}$ \\
    \hline
        0&0&0   &   0&0&0&0&0&0&0&0&0&0  \\
        0&0&1   &   0&0&0&0&0&0&0&0&0&0  \\
        0&1&0   &   0&0&0&0&1&0&0&0&0&0  \\
        0&1&1   &   0&1&1&0&1&1&0&1&0&1  \\
        1&0&0   &   0&0&0&0&1&0&0&0&0&0  \\
        1&0&1   &   0&0&0&0&1&0&0&0&0&0  \\
        1&1&0   &   1&0&1&0&1&1&0&1&0&1  \\
        1&1&1   &   1&1&1&1&1&1&1&0&1&1  \\
    \hline
    \end{tabular}
\end{table}

We observe that $n_3$, $n_6$ and $n_{10}$ are functionally equivalent, as well as $n_4$, $n_7$, and $n_{10}$. 
Therefore, these three circuits are design choices that a logic synthesis algorithm, such as resubstitution introduced in \Cref{subsec:resub}, may choose from. 

Most state-of-the-art logic synthesis algorithms optimize for AIG size or depth, resulting in a tie between $N_A$ and $N_B$ while deeming $N_C$ suboptimal, as shown in \Cref{tab:example-cost-evaluation}. However, depending on the target application, there may be very different conclusions on which network is the best. For example, if the target technology imposes a path-balancing constraint, the networks $N_A$ and $N_C$ are preferred over $N_B$ as they have lower maximum skew; when the (structural) multiplicative complexity is of concern, the XOR nodes ($n_9$ and $n_{10}$) are free, and the network $N_C$ has the lowest cost.

\begin{table}[h]
\caption{\small Evaluations of different cost functions on networks $N_A$, $N_B$, and $N_C$. Size evaluation of a network sums up the node costs, and depth evaluation finds the maximum costs. We highlight the optimal candidate(s) among $N_A$, $N_B$, and $N_C$, which has the lowest cost, under each cost definition.}\label{tab:example-cost-evaluation}
\small
\tabcolsep = 2pt
    \centering
    \begin{tabular}{|l|llll n |llll n |lllll n |}
   \hline
       \rowcolor{lightgray} 
       & \multicolumn{5}{c|}{Network $N_A$} &\multicolumn{5}{c|}{Network $N_B$} & \multicolumn{6}{c|}{Network $N_C$}  \\
    \cline{2-17}
       \rowcolor{lightgray} 
       \multirow{-2}{*}{Costs}  
       & $n_1$ & $n_2$ & $n_3$ & $n_4$ & $A$ & $n_2$ & $n_5$ & $n_6$ & $n_7$ & $B$ & $n_1$ & $n_2$ & $n_8$ & $n_9$ & $n_{10}$ & $C$ \\ 
    \hline
        AIG size  & 1 & 1 & 1 & 1 & \opt{4} & 1 & 1 & 1 & 1 &  \opt{4} & 1 & 1 & 1 & 3 & 3 & 9 \\
        AIG depth & 1 & 1 & 2 & 2 & \opt{2} & 1 & 1 & 1 & 1 &  \opt{2} & 1 & 1 & 3 & 2 & 5 & 5 \\
    \hline
        Max. skew & 0 & 0 & 0 & 0 & \opt{0} & 0 & 0 & 1 & 1 & 1 & 0 & 0 & 0 & 0 & 0 & \opt{0} \\
        MC~\cite{TestaSAM19} & 1 & 1 & 1 & 1 & 4 & 1 & 1 & 1 & 1 & 4 & 1 & 1 & 1 & 0 & 0 & \opt{3} \\
    \hline
    \end{tabular}
\end{table}

This example shows that re-applying existing AIG size- or depth-oriented algorithms on emerging technologies and applications may be suboptimal, as they may make decisions that increase the actual cost of concern. In contrast, dedicated specialized algorithms for various applications are more effective as they take more accurate cost metrics into account, even if these algorithms still work on technology-independent representations to keep themselves scalable. However, as cost evaluation is often an estimation, there could be many ways to define the cost functions. Whenever a new cost function arises, weeks to months of engineering effort are needed to develop a specialized algorithm. Thus, a cost-generic resynthesis algorithm to quickly test the effectiveness of each candidate cost function is helpful to speed up prototyping and reduce development time.

\section{Customizable Cost Functions}\label{sec:definition}
Comparing the costs of local optimization choices is essential to any logic optimization algorithm. The key contribution of this work is to generalize the cost evaluation and unify the cost definition interface. Let the \emph{global cost} $\Gamma \in \mathbb{Z}_{\geq 0}$ of a network be the cost evaluation result according to a given cost function. For the sake of scalability, optimization problems are localized within a window. Therefore, we need to evaluate or predict the influence of each local optimization choice on the global cost. 


However, local evaluation results do not always relate directly to global costs. For example, when the cost function is circuit depth, the levels of the window outputs depend on the input levels. Besides, local optimization does not always imply global optimization. In this example, reducing the depths of windows does not necessarily improve the network depth if the window is not on the critical path or if there are multiple critical paths. To this end, in this section, we introduce two mechanisms in our definition system, \emph{context propagation} and \emph{independent node contribution}, to associate local cost evaluation with global evaluation.


\subsection{Context Propagation}\label{subsec:context}
We define the \emph{context} of a node $n$, denoted by $\gamma_n$, as the information involved in cost evaluation that cannot be determined with $n$ alone. The computation of the context is defined by the \emph{context propagation function} $\Phi^\gamma$. The context propagation function specifies the required information outside the window and expresses how this information can be derived from the network. To improve the propagation efficiency, we restrict the context propagation function to a recursive function of node fanins, i.e., $\gamma_{n} = \Phi^\gamma(n, I^\gamma_n)$, where $I^\gamma_n = \{\gamma_i:i\in \delta^-(n)\}$. In other words, the context evaluation of a node $n$ can only depend on the contexts of its fanins. As a result, cost functions involving complicated global calculations outside the TFI cone, such as the sum of all-pairs-min-cut~\cite{SAPMC}, cannot be accurately expressed in our framework and must be approximated. 

\begin{example}
    Utilizing the context propagation function, we can evaluate the number of \emph{reconvergence} in a given network, representing the number of node pairs that are connected with at least two distinct paths. For simplicity, we assume the given network contains only two input nodes. During the evaluation, we store the context, $\gamma_n$, for node $n$ as the set of nodes in $n$'s TFI. Take $N_C$ in \Cref{fig:example-network-3} as an example. Primary inputs $a$, $b$, $c$ have TFIs as sets of themselves: $\gamma_a = \{a\}$, $\gamma_b = \{b\}$, and $\gamma_c = \{c\}$. Node $n_1$ stores the context $\gamma_1 = \{a,c, n_1\}$, which can be propagated using the fanin contexts $I^\gamma_1 = \{\gamma_a, \gamma_c\}$:
    \begin{equation*}
        \gamma_1 = \Phi^\gamma(n, I^\gamma_n) = \gamma_a \cup \gamma_c \cup \{n_1\},
    \end{equation*}
    which expresses the union of fanin contexts and the nodes $n_1$. Similarly, $\gamma_2 = \gamma_b \cup \gamma_c \cup \{n_2\} = \{b, c, n_2\}$. Then, $\gamma_1$ and $\gamma_2$ can be utilized to check the number of reconvergences (\#reconv) with one endpoint at $n_8$. 
    \begin{equation*}
        \text{\#reconv}(n_8) = |\gamma_1 \cap \gamma_2|,
    \end{equation*}
    which is the cardinality of $\gamma_1$ and $\gamma_2$'s intersection. This equation correctly derives the convergence, as each node $n'$ occurs in $n$'s both fanin TFI implies a unique reconvergence between $n'$ and $n$. We can sum up the reconvergences at each node to evaluate the total reconvergence of the entire network. This example demonstrates that our definition system applies to complicated cost definitions.
\end{example}

\subsection{Independent Node Contribution}
We evaluate the global cost function using the \emph{node contribution function}, denoted by $\Phi^\Gamma$. The node contribution function specifies how the context of each node affects the global cost $\Gamma$. More specifically, $\Phi^\Gamma$ takes the context of $n$ and output the updated the global cost, i.e., $\Gamma' = \Phi^\Gamma(\Gamma, \gamma_n)$. 

Consequently, the cost generated at each node contributes directly and independently to the cost evaluation of the entire network. As illustrated in \Cref{tab:example-cost-evaluation}, the evaluation of size cost and MC cost collects the individual contributions by adding them to the final cost, e.g., $\Gamma\!=\!\Gamma + 1$ for each AND node. For the depth cost and the maximum skew evaluation, we acquire the nodes' contribution from the context $\gamma_n$, and collect them to the total cost by $\Gamma\!=\!\max(\Gamma, \gamma_n)$. Note that this feature allows us to evaluate the cost caused by a local substitution without considering the nodes outside the window. Indeed, to evaluate a local optimization choice, we first calculate the cost saved by applying $\Phi^\Gamma$ to each node in the MFFC, and we evaluate the cost of the substitution candidate by applying $\Phi^\Gamma$ again to nodes in the dependency circuit. Substitution reduces the global cost if the net cost change is negative. 

\subsection{Flexibility and Evaluation Efficiency Analysis}
\begin{table}[tb]
\caption{\small Cost Function Definition Examples}
\centering
\def\arraystretch{1.5}\tabcolsep 2pt
\def\thefootnote{a}\footnotesize
\begin{tabular}{l|l|l}
\hline
 & Cost Name & Cost Functions \\
\hline
\multirow{2}{*}{$\Phi_1$} & \multirow{2}{*}{MC~\cite{TestaSAM19}} & 
    $\Phi^\Gamma_1(\Gamma, \gamma_n)=
\begin{cases}
\Gamma +1& n\text{ is AND}\\
\Gamma& \text{otherwise}
\end{cases}$ \\[1ex]
& &  
    $\Phi^\gamma_1(n, I^\gamma_n) = \varnothing$ \\[3ex]
\multirow{2}{*}{$\Phi_2$} & \multirow{2}{*}{T depth\cite{haner2020lowering}} & 
    $\Phi^\Gamma_2(\Gamma, \gamma_n)=\max (\Gamma, \gamma_n)$ \\[1ex]
& & 
    $\Phi^\gamma_2(n, I^\gamma_n)= 
\begin{cases}
\max\limits_{\gamma_i\in I^\gamma_n}\gamma_{i} + 1 & n\text{ is AND}\\
\max\limits_{\gamma_i\in I^\gamma_n}\gamma_{i} & \text{otherwise} \\
\end{cases}$ \\[5ex]
\multirow{2}{*}{$\Phi_3$} & \multirow{2}{*}{Support reduction~\cite{Support}} & 
    $\Phi^\Gamma_3(\Gamma, \gamma_n)=\Gamma + |\gamma_n|$ \\[1ex]
& &  
    $\Phi^\gamma_3(n, I^\gamma_n)=
\begin{cases}
\{n\} & n\in\text{PI} \\
\bigcup\limits_{\gamma_i\in I^\gamma_n}\gamma_{i} & \text{otherwise}
\end{cases}$ \\[4ex]
\hline
\end{tabular}
\label{tab:cost-example}
\end{table}
Users can specialize their costs by customizing context propagation and node contribution functions. \Cref{tab:cost-example} gives three additional examples of practical cost function definitions within our system. $\Phi_1$ defines multiplicative complexity, where each node contributes one unit to $\Gamma$ if it is an AND node. For T depth~\cite{haner2020lowering} ($\Phi_2$), the node levels are recursively derived using context propagation, and the maximum node level determines $\Gamma$. $\Phi_3$ reduces the sum of supports in the network. The support set is stored as the context and is calculated with the union of fanin supports. These examples demonstrate that the cost function in our framework is highly customizable with both parameter and model modifications.

Cost evaluation propagates context and collects node contribution in topological order. This way, the fanins of a node $n$ are evaluated and have their contexts updated before $n$, and the updated context remains correct in later optimization because the TFI cone is traversed and fixed. Additionally, we evaluate each node exactly once. Therefore, the overhead we introduce to be generic does not affect the scalability as the complexity grows linearly with the network size.

\section{Cost-Generic Resynthesis}\label{sec:algorithm}
\begin{figure}[t]
    \centering
    \includegraphics[width=.98\linewidth]{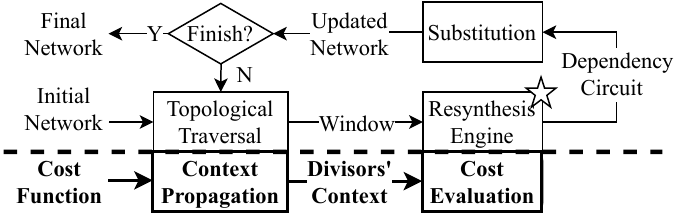}
    \centering
    \caption{\small AnySyn: A cost-generic logic synthesis framework.}
    \label{fig:flow}
\end{figure}

\begin{table}[b]
\centering
\def\arraystretch{1}\tabcolsep 1pt
\def\thefootnote{*}\footnotesize
\begin{minipage}{\linewidth}
\caption{\small Dependency circuit structures.}\label{tab:dependency-circuit}
\begin{tabular}{c|c|c|c|c|c}
    \toprule
    \includegraphics[width=.125\linewidth]{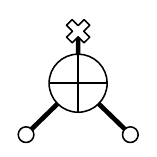} &
    \includegraphics[width=.125\linewidth]{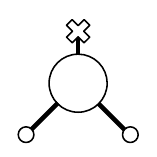} &
    \includegraphics[width=.125\linewidth]{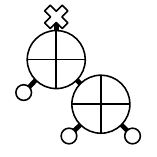} &
    \includegraphics[width=.125\linewidth]{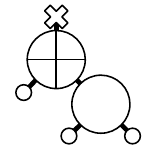} &
    \includegraphics[width=.125\linewidth]{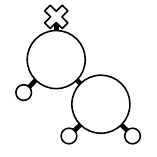} &
    \includegraphics[width=.125\linewidth]{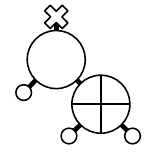} \\
    $\Theta(d)$ & 
    $\mathcal{O}(d^2)$ & 
    $\Theta(d^2)$ & 
    $\Theta(d^2)$ & 
    $\mathcal{O}(d^3)$ &
    $\mathcal{O}(d^3)$ \\ \midrule
    \includegraphics[width=.15\linewidth]{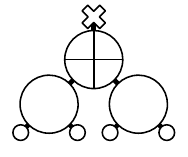} &
    \includegraphics[width=.15\linewidth]{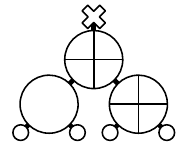} &
    \includegraphics[width=.15\linewidth]{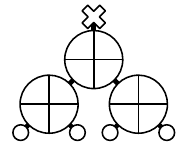} &    
    \includegraphics[width=.15\linewidth]{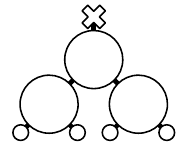} &
    \includegraphics[width=.15\linewidth]{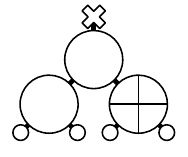} &
    \includegraphics[width=.15\linewidth]{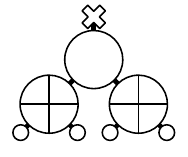}
 \\
    $\Theta(d^2)$ & 
    $\Theta(d^2)$ & 
    $\Theta(d^2)$ &
    $\mathcal{O}(d^4)$ & 
    $\mathcal{O}(d^4)$ & 
    $\mathcal{O}(d^4)$ \\ \bottomrule
    \multicolumn{6}{l}{\scriptsize $\Bigxor$ represents XOR node and $\Bigcircle$ represents AND node.}\\
\end{tabular}
\end{minipage}
\end{table}

Our cost-generic logic optimization framework is based on resubstitution~\cite{Mishchenko06}, but adds cost-dependent ingredients, including the customizable cost function as input to the framework, context propagation, and cost evaluation in the cost-generic resynthesis algorithm. The workflow is shown in \Cref{fig:flow}.  In this section, we introduce the cost-generic resynthesis engine with two heuristics to improve its efficiency and cost-generality. Although the concept of customizable cost functions, their evaluation, and cost-generic resynthesis can be applied to any logic network representation, our implementation uses XAGs as the underlying data structure to facilitate a more flexible cost function definition.

\subsection{Cost-Generic Resynthesis Engine}
Similar to classic resubstitution algorithms~\cite{Mishchenko06,RienerTASM18,LeeRMBM22}, we first traverse the network in topological order and extract one window for each node as an optimization target. Given a resynthesis problem~\cite{LeeM23} consisting of a target node (function) and a set of divisors, we enumerate all possible dependency circuit structures and all divisor combinations at the input of the dependency circuit. For each combination, we check if the output function is equivalent to the target. \Cref{tab:dependency-circuit} depicts all the dependency circuit structures associated with the time complexity to enumerate all divisor combinations.

Unlike resubstitution, which returns the first functional equivalent circuit, our algorithm constructs a \emph{solution forest} that collects all the candidates, evaluates the cost of each, and returns the one with the lowest cost. Then, we substitute the window, update the network, and repeat the optimization for each node in the network. 


\begin{figure}[bt]
    \centering
    \includegraphics[width=.7\linewidth]{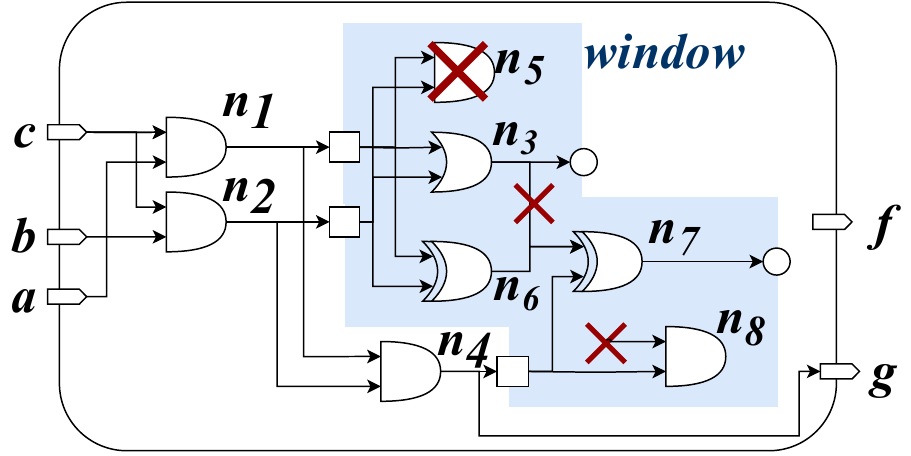}
    \centering
    \caption{\small Example of a cost-generic resynthesis problem. $n_1$, $n_2$ and $n_4$ are divisors, and $f$ is the target node. Nodes $n_3$ and $n_7$ are functionally equivalent to $f$, therefore, are substitution candidates.}
    \label{fig:resyn-example}
\end{figure}

\subsection{Two Techniques to Prune Dependency Circuits}
Many specialized algorithms that are carefully designed for a specific cost function benefit from some cost-based pruning techniques such as step-by-step $k$-resubstitution~\cite{brayton2006scalable} and tree-balancing~\cite{mishchenko2011delay}. As a generic algorithm, some run-time overhead is expected compared to specialized algorithms.
To mitigate this, we propose two search-space pruning techniques based on functionality and structural information. 

The examples are illustrated in \Cref{fig:resyn-example}. Besides $n_1$ and $n_2$, we also collect $n_4$ as divisors. After enumerating all the dependency circuit structures, we find that both $n_3$ and $n_7$ are functionally equivalent to $f$. The shaded area represents the solution forest generated for the window, where circles represent the inputs and outputs. All outputs are substitution candidates for $f$. Red crosses show the pruned structures during the search.

\textbf{Technique 1:} pruning based on Boolean properties in the XAG. For AND nodes, $(f=x\wedge y) \Rightarrow (f\Rightarrow x)$. Thus, $\neg(f\Rightarrow x) \Rightarrow \forall y: f\neq xy$. For example, $n_8\neq f$ because $f=n_1\vee n_2\nRightarrow n_1\wedge n_2$, and we prune $n_8$ without enumerating the other input of the AND node. For XORs, $(f=x\oplus y) \Leftrightarrow (x=f\oplus y)$. Therefore, we store $S = \{f\oplus y\,|\,y\in \text{Divisors}\}$ as a hash table and prune the infeasible nodes efficiently by a hash table lookup, i.e., $\{x\,|\,x\notin S\}$. For example, we assert $n_3$ cannot be an input to $n_7$ because $n_7$ takes a divisor $n_4$ as input, and the functionality of $n_3$ does not exist in the hash table. 

\textbf{Technique 2:} pruning based on structural equivalence. We prune a dependency circuit if it introduces redundant nodes to the window. For example, we prune $n_5$, which requires an AND of $n_1$ and $n_2$, because the structurally equivalent node $n_4$ exists and is a divisor. This technique is implemented by merging all the dependency circuits into a structural hashed logic network. Apply structural hashing disables optimizations that require node duplication or buffer insertion. However, without this technique, the search engine would find abundant redundant circuits that recreate existing divisors.

\subsection{Deterministic Dependency Circuit Construction}
Although we reduce the run-time with the pruning techniques, the complexity grows exponentially with the circuit size. As a result, searching large dependency circuits is time-consuming. This intrinsic feature of resubstitution prioritizes size optimization. To mitigate run-time overhead and bias, we use deterministic SOP and ESOP decompositions to add larger circuits to the solution forest. SOP and ESOP comprise two-level multi-input AND, OR, and XOR nodes, using only the window's input. When decomposing the multi-input nodes into two-input XAG nodes, users can optionally define a partial order for the context, and our algorithm will sort the inputs in ascending order. Overall, decomposition is efficient and improves the circuit structure's generality.

\section{Experimental Results}\label{sec:result}
The proposed framework is implemented as a new feature in an open-source project.\footnote{https://github.com/lsils/mockturtle.} In this section, we illustrate the effectiveness of AnySyn as a cost-generic algorithm on various cost definitions and discuss the run-time overhead using experimental results.

\subsection{Optimization of Various Cost Functions}\label{subsec:exp-costs}
This experiment demonstrates the effectiveness of AnySyn on nine different cost functions. The EPFL benchmark suite~\cite{EPFLBmark} is used and preprocessed with two iterations of the \texttt{compress2rs} script in ABC~\cite{abc} to eliminate trivial redundancies. We compare the result of optimizing corresponding cost functions against two baselines, ``size opt.'' and ``depth opt.''. Both AnySyn and two baselines apply the same cost-generic resynthesis algorithm introduced in \Cref{sec:algorithm} with the same window sizes and number of divisors but optimize for different cost functions. AnySyn selects the substitution candidates according to the specified cost function. Two baselines, ``size opt.'' and ``depth opt.'', select size-optimal and depth-optimal candidates, thus, are \emph{non-specialized}.
\begin{table}[b]
\caption{\small Optimization results of various cost functions.}\label{tab:all-cost}
\small
\centering
\def\thefootnote{a}\footnotesize
\begin{minipage}{\linewidth}    
\centering
\begin{tabular}{l|rrrr}
\hline
\multicolumn{5}{c}{\textbf{Size-like cost functions}}                   \\
\hline
 & \multicolumn{1}{c}{initial} & \multicolumn{1}{c}{size opt.} & \multicolumn{1}{c}{depth opt.} & \multicolumn{1}{c}{\textbf{AnySyn}} \\
\hline
XAG size$^{\scriptsize a}$      & 3359   & 3329   & 4245  & \textbf{2948}  \\
MC~\cite{TestaSAM19}            & 3356   & 2538   & 4025  & \textbf{1942}  \\
Total skew$^{\scriptsize b}$    & 104788 & 89212  & 81669 & \textbf{62781} \\
Reconv$^{\scriptsize c}$           & 21599  & 15984  & 36380 & \textbf{14247} \\
FFLC\cite{Calvino23}          & 4530   & 3902   & 6280  & \textbf{3844}  \\
\hline
Geomean       & 10294  & 8599   & 12610 & \textbf{7225}  \\
Ratio         & 1.00   & 0.84   & 1.22  & \textbf{0.70} \\
\hline
\hline
\multicolumn{5}{c}{\textbf{Depth-like cost functions}}                  \\
\hline
 & \multicolumn{1}{c}{initial} & \multicolumn{1}{c}{size opt.} & \multicolumn{1}{c}{depth opt.} & \multicolumn{1}{c}{\textbf{AnySyn}} \\
\hline
XAG depth$^{\scriptsize a}$     & 116.57 & 117.05 & 85.41 & \textbf{59.12} \\
T depth\cite{haner2020lowering}      & 116.57 & 110.45 & 52.70 & \textbf{43.71} \\
Max skew$^{\scriptsize d}$      & 111.34 & 111.00 & 56.58 & \textbf{55.51} \\
AND chain$^{\scriptsize e}$     & 116.57 & 103.94 & 27.45 & \textbf{5.79}  \\
\hline
Geomean       & 115.24 & 110.51 & 51.42 & \textbf{30.19} \\
Ratio         & 1.00   & 0.96   & 0.45  & \textbf{0.26}  \\
\hline
\end{tabular}
\footnotetext[1]{\small In these two experiments the baselines are AIG (instead of XAG) size or depth optimization and over-estimate the cost of XOR nodes.}
\footnotetext[2]{\small The sum of the fanins' level difference at each node.}
\footnotetext[3]{\small The number of reconvergence in the network.}
\footnotetext[4]{\small The maximum of the fanins' level difference among all the nodes.}
\footnotetext[5]{\small The length of longest AND chain, i.e., consecutive AND nodes.}
\end{minipage}
\end{table}

The results are shown in \Cref{tab:all-cost}. All listed cost functions can be defined within ten lines of \texttt{C++} code. Based on their node contribution characteristics, we categorize cost functions into size-like and depth-like. Size-like functions add nodes' contributions to the global cost, whereas depth-like functions return the maximum of nodes' contributions. Entries in the table are geometric means of the cost values among all benchmarks. 

AnySyn achieves the best quality of results on all cost functions. On average, AnySyn outperforms existing cost-generic algorithms by $14\%$ and $19\%$. This experiment demonstrates the disadvantages of simple AIG/XAG size and depth-oriented optimizations and the benefit of having a more accurate cost evaluation in the resynthesis problem. Moreover, the success in both size- and depth-like costs shows that AnySyn generalizes the two seemingly orthogonal optimization philosophies in a unified framework. 
\begin{table}[tb]
\caption{\small FFLC optimization results}\label{tab:fflc}
\def\thefootnote{a}\footnotesize
\begin{minipage}{\linewidth} 
\centering
\begin{tabular}{l|rrrr}
\hline
 & \multicolumn{1}{c}{initial} &baseline$^{\scriptsize a}$ & special.$^{\scriptsize b}$ & \textbf{ours} \\
\hline
AC97 controller    & 13039  & \textbf{12971} & 12979          & 13018           \\
AES core     & 24738  & 24511          & \textbf{23972} & 24374           \\
DES area     & 5177   & 5162           & \textbf{5093}  & 5145            \\
DES perf     & 99730  & 99000          & \textbf{95820} & 98572           \\
DMA          & 26614  & 26589          & 25706          & \textbf{24366}  \\
DSP          & 47734  & 47187          & \textbf{46561} & 46700           \\
Ethernet     & 69226  & 69160          & 69011          & \textbf{68953}  \\
I2C Controller         & 1136   & 1124           & \textbf{1087}  & 1102            \\
Memory Controller     & 10410  & 10303          & \textbf{10150} & 10198           \\
PCI Bridge32 & 20628  & 20616          & \textbf{20477} & 20490           \\
RISC         & 75005  & 74631          & 73260          & \textbf{73168}  \\
SAS Controller         & 729    & 729            & \textbf{722}   & 726             \\
MC68HC11E SPI   & 961    & 960            & \textbf{950}   & 952             \\
SPI IP          & 3797   & 3736           & 3692           & \textbf{3587}   \\
Single Slot PCM       & 497    & 497            & 497            & 497             \\
SystemC DES   & 11542  & 11444          & 11157          & \textbf{11111}  \\
SystemC AES   & 3252   & 3231           & \textbf{3061}  & 3111            \\
TV80 Processor         & 8603   & 8507           & \textbf{8158}  & 8293            \\
USB Function    & 16145  & 16059          & 15986          & \textbf{15870}  \\
USB 1.1 PHY      & 524    & 524            & \textbf{512}   & 516             \\
VGA/LCD      & 112806 & 112802         & 112747         & \textbf{112533} \\
Conmax    & 43041  & 42338          & \textbf{41476} & 42102           \\
\hline
Ratio        & 1.0000 & 0.9941 & 0.9765 & 0.9785 \\
\hline
\end{tabular}
\footnotetext[1]{\small We use \texttt{compress2rs} as baseline.}
\footnotetext[2]{\small The specialized algorithm is the state-of-the-art FFLC optimization algorithm\cite{Calvino23}.}
\end{minipage}
\end{table}

\subsection{Comparison with a Specialized Algorithm}\label{subsec:exp-stoa}
We compare our algorithm to a recently proposed specialized algorithm optimizing for the \emph{factored-form literal count}~(FFLC)~\cite{Calvino23} and show that AnySyn achieves comparable optimization quality. The cost function is defined as $\text{FFLC} = |O|+2\times |G|-|M|$, where $O$, $G$, and $M$ are the set of primary output, gates, and multiple fanout gates, respectively. The context propagation and node contribution functions are written as \Cref{eqn:fflc}. 
The term $|O|$ is neglected because it remains constant throughout the optimization. 
\begin{equation}\label{eqn:fflc}
\Phi^\Gamma(\Gamma, \gamma_n)= \Gamma + \gamma_n \,,\,
\Phi^\gamma(n, I_\gamma)= 
\begin{cases}
2, & |\delta^+(n)| > 1\\
1, & \text{otherwise} \\
\end{cases}.
\end{equation}

We use the same selected benchmarks as~\cite{Calvino23} from the IWLS'05 benchmark suite\footnote{Available: \url{iwls.org/iwls2005/benchmarks.html}}, preprocessed with two runs of \texttt{compress2rs}. For a fair comparison, we integrate our method into ABC's \texttt{compress2rs} flow by replacing all resubstitution calls with AnySyn using the same window sizes. The baseline is two more runs of \texttt{compress2rs}. 

\Cref{tab:fflc} shows the comparison results. The best results are highlighted in bold. The optimization quality of AnySyn is comparable to the specialized algorithm and even outperforms it on some benchmarks. Notice that the non-specialized algorithm achieves the best result on the benchmark ``AC97 controller'', showing that greedily choosing the best local optimization choices does not always result in the lowest global cost. However, in general, we optimize the FFLC by $2.15$\%, which is close to $2.35$\% by the specialized algorithm, while the non-specialized algorithm reduces the FFLC by only $0.06$\%.

\subsection{Scalability Analysis}
As a more generic algorithm, AnySyn is expected to have some run-time overhead over specialized algorithms. In this section, we analyze such overhead and show that AnySyn is still scalable to larger benchmarks. 

\begin{figure}[t]
    \centering
    \includegraphics[width=.8\linewidth]{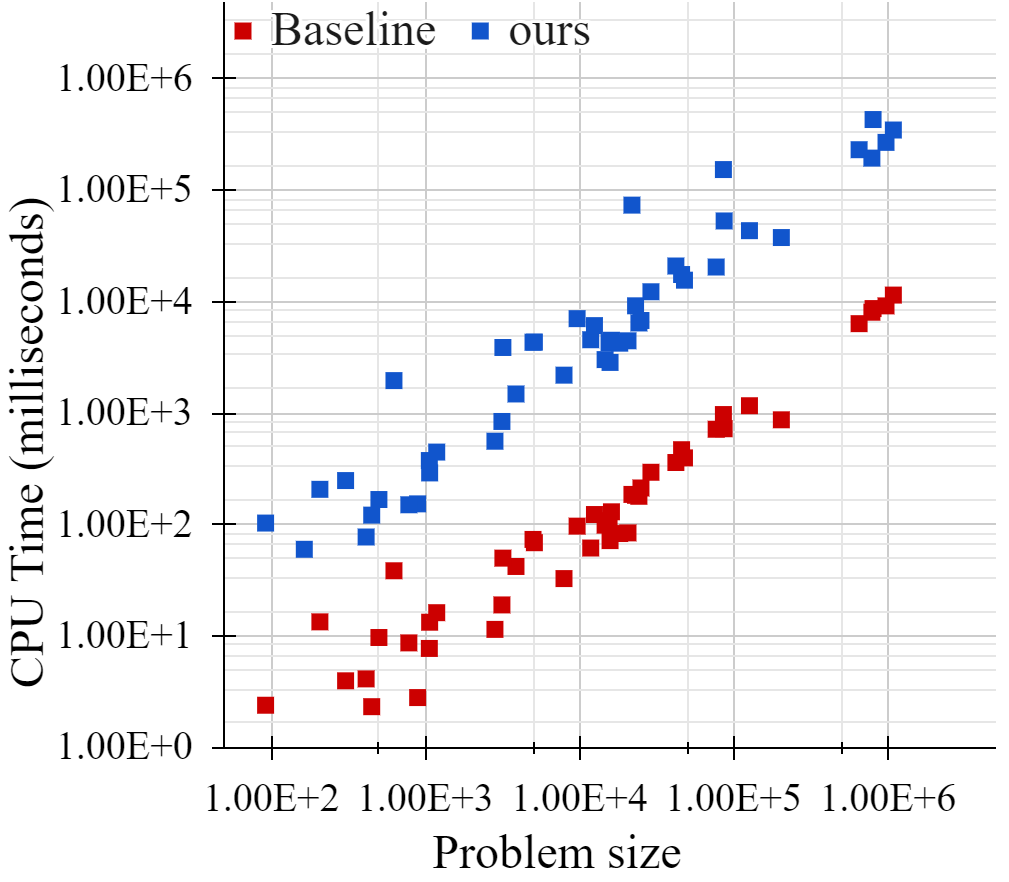}
    \caption{\small CPU time comparison. Each data point represents a run of AnySyn on one benchmark. Problem size is the number of nodes in the benchmark logic network. We record the CPU time of the entire execution including network traversal, window construction, and solving resynthesis problems.}
    \label{fig:runtime-exp}
\end{figure}

\Cref{fig:runtime-exp} shows the scalability of AnySyn. The $x$-axis is the benchmark size, and the $y$-axis is the CPU time in milliseconds to resynthesize the network and optimize each node once. The baseline is a scalable resubstitution algorithm~\cite{WindowResyn}, and the IWLS'05 and EPFL benchmark suites are used. Although our implementation is slower, the run-time overhead is bounded by a constant. The CPU times of both algorithms grow linearly with the problem size on the double-logarithm graph, showing that they have the same asymptotic behavior. The network sizes range from $100$ to $1$ million, and our algorithm takes less than $6$ minutes on the largest benchmark.  

\section{Conclusions}
In this paper, we show the benefits of modeling technology-dependent cost functions in technology-independent logic networks and propose a cost-generic logic synthesis framework. AnySyn eases experimenting with new cost functions, requiring less than ten lines of code to define a new cost function and getting a specialized optimization for the target application.  The proposed cost definition method is flexible and compatible with various optimization objectives. Moreover, with the help of divisors and more accurate cost estimation, our cost-generic resynthesis can find the appropriate dependency circuit to optimize the total cost. Experiments show that AnySyn achieves similar better results than specialized algorithms.

\bibliographystyle{IEEEtran}
\bibliography{main}

\begin{thebibliography}{10}
\providecommand{\url}[1]{#1}
\csname url@samestyle\endcsname
\providecommand{\newblock}{\relax}
\providecommand{\bibinfo}[2]{#2}
\providecommand{\BIBentrySTDinterwordspacing}{\spaceskip=0pt\relax}
\providecommand{\BIBentryALTinterwordstretchfactor}{4}
\providecommand{\BIBentryALTinterwordspacing}{\spaceskip=\fontdimen2\font plus
\BIBentryALTinterwordstretchfactor\fontdimen3\font minus \fontdimen4\font\relax}
\providecommand{\BIBforeignlanguage}[2]{{%
\expandafter\ifx\csname l@#1\endcsname\relax
\typeout{** WARNING: IEEEtran.bst: No hyphenation pattern has been}%
\typeout{** loaded for the language `#1'. Using the pattern for}%
\typeout{** the default language instead.}%
\else
\language=\csname l@#1\endcsname
\fi
#2}}
\providecommand{\BIBdecl}{\relax}
\BIBdecl

\bibitem{Mishchenko06}
A.~Mishchenko and R.~K. Brayton, ``Scalable logic synthesis using a simple circuit structure,'' in \emph{Proceedings of IWLS}, 2006.

\bibitem{TestaSAM19}
E.~Testa, M.~Soeken, L.~G. Amar{\`{u}}, and G.~De~Micheli, ``Reducing the multiplicative complexity in logic networks for cryptography and security applications,'' in \emph{{DAC} 2019}, p.~74.

\bibitem{meuli2022xor}
G.~Meuli, M.~Soeken, and G.~De~Micheli, ``{Xor-And-Inverter} graphs for quantum compilation,'' \emph{npj Quantum Information}, vol.~8, no.~1, pp. 1--11, 2022.

\bibitem{Schnorr88}
C.~Schnorr, ``The multiplicative complexity of boolean functions,'' in \emph{{AAECC} 1988}, vol. 357, pp. 45--58.

\bibitem{Calvino23}
A.~Tempia~Calvino, A.~Mishchenko, H.~Schmit, E.~Mahintorabi, G.~De~Micheli, and X.~Xu, ``Improving standard-cell design flow using factored form optimization,'' in \emph{Proceedings of {DAC}}, 2023.

\bibitem{KuehlmannPKG02AIG}
A.~Kuehlmann, V.~Paruthi, F.~Krohm, and M.~K. Ganai, ``Robust {Boolean} reasoning for equivalence checking and functional property verification,'' \emph{{IEEE} Trans. Comput. Aided Des. Integr. Circuits Syst.}, vol.~21, no.~12, pp. 1377--1394, 2002.

\bibitem{HalecekFS17XAG}
I.~H{\'{a}}lecek, P.~Fiser, and J.~Schmidt, ``Are {XORs} in logic synthesis really necessary?'' in \emph{{DDECS} 2017}, pp. 134--139.

\bibitem{MishchenkoCB06}
A.~Mishchenko, S.~Chatterjee, and R.~K. Brayton, ``{DAG}-aware {AIG} rewriting: {A} fresh look at combinational logic synthesis,'' in \emph{{DAC} 2006}, pp. 532--535.

\bibitem{RienerTASM18}
H.~Riener, E.~Testa, L.~G. Amar{\`{u}}, M.~Soeken, and G.~De~Micheli, ``Size optimization of {MIGs} with an application to {QCA} and {STMG} technologies,'' in \emph{{NANOARCH} 2018}, 2018, pp. 157--162.

\bibitem{LeeM23}
S.-Y. Lee and G.~De~Micheli, ``Heuristic logic resynthesis algorithms at the core of peephole optimization,'' \emph{{IEEE} Trans. Comput. Aided Des. Integr. Circuits Syst.}, 2023.

\bibitem{SAPMC}
P.~Kudva, A.~Sullivan, and W.~Dougherty, ``Metrics for structural logic synthesis,'' in \emph{{ICCAD} 2002}, pp. 551--556.

\bibitem{haner2020lowering}
T.~H{\"a}ner and M.~Soeken, ``Lowering the t-depth of quantum circuits by reducing the multiplicative depth of logic networks,'' \emph{arXiv preprint arXiv:2006.03845}, 2020.

\bibitem{Support}
L.~Machado and J.~Cortadella, ``Support-reducing decomposition for fpga mapping,'' \emph{{IEEE} Trans. Comput. Aided Des. Integr. Circuits Syst.}, vol.~39, no.~1, pp. 213--224, 2020.

\bibitem{LeeRMBM22}
S.-Y. Lee, H.~Riener, A.~Mishchenko, R.~K. Brayton, and G.~De~Micheli, ``A simulation-guided paradigm for logic synthesis and verification,'' \emph{{IEEE} Trans. Comput. Aided Des. Integr. Circuits Syst.}, vol.~41, no.~8, pp. 2573--2586, 2022.

\bibitem{brayton2006scalable}
A.~M.~R. Brayton, ``Scalable logic synthesis using a simple circuit structure,'' in \emph{Proc. IWLS}, vol.~6, 2006, pp. 15--22.

\bibitem{mishchenko2011delay}
A.~Mishchenko, R.~Brayton, S.~Jang, and V.~Kravets, ``Delay optimization using {SOP} balancing,'' in \emph{{ICCAD} 2011}, pp. 375--382.

\bibitem{EPFLBmark}
L.~G. Amar{\`u}, P.-E. Gaillardon, and G.~De~Micheli, ``The epfl combinational benchmark suite,'' in \emph{{IWLS} 2015}.

\bibitem{abc}
R.~Brayton and A.~Mishchenko, ``{ABC: An Academic Industrial-Strength Verification Tool},'' in \emph{Proceedings of {CAV}}, 2010, pp. 24--40.

\bibitem{WindowResyn}
H.~Riener, S.-Y. Lee, A.~Mishchenko, and G.~De~Micheli, ``Boolean rewriting strikes back: Reconvergence-driven windowing meets resynthesis,'' in \emph{{ASP-DAC} 2022}, pp. 395--402.

\end{thebibliography}

\end{document}